\documentclass[aps,prl,floatfix,twocolumn,showpacs,groupedaddress]{revtex4} 
\usepackage{graphicx,color}%
\newcommand{\be}{\begin{equation}}
\newcommand{\ee}{\end{equation}}
\newcommand{\bea}{\begin{eqnarray}}
\newcommand{\eea}{\end{eqnarray}}

\newcommand{\p}{\partial}

\newcommand{\rd}{{\rm d}}

\newcommand{\EQ}{\begin{equation}}
\newcommand{\EN}{\end{equation}}

\begin{document}
\title{On the Mass Spectrum of the Two--dimensional 
$O(3)$ Sigma Model with $\theta$ Term}
\author{D. Controzzi $^a$\footnote{On leave from: {\em International School for Advanced Studies,
Trieste, Italy.} } and G. Mussardo $^b$}
\affiliation{ $^a$ Institute for Theoretical Physics
Valckenierstraat 65, 1018XE Amsterdam, The Netherlands \\
$^b$ International School for Advanced Studies and INFN,
Via Beirut 1, 34100 Trieste, Italy}

\begin{abstract}
\par
Form Factor Perturbation Theory is applied to study 
the spectrum of the $O(3)$ non--linear sigma model with the topological 
term in the vicinity of $\theta = \pi$. Its effective action 
near this value is given by the non--integrable double Sine--Gordon 
model. Using previous results by Affleck and the explicit 
expressions of the Form Factors of the exponential operators 
$e^{\pm i\sqrt{8\pi} \varphi(x)}$, we show that the spectrum consists 
of a stable triplet of massive particles for all values of $\theta$ 
and a singlet state of higher mass. The singlet is a stable 
particle only in an interval of values of $\theta$ close to $\pi$ 
whereas it becomes a resonance below a critical value $\theta_c$. 
\end{abstract}
\maketitle

The $O(3)$ non--linear sigma model is a two--dimensional quantum field 
theory  
for a 3-component, unit-vector field $n_\alpha$ ($\alpha=1,2,3 \; ; 
n_\alpha^2=1$), with the Euclidean action given by 
\be
{\cal A} _{\theta} \,=\, \frac{1}{2f^2} \int \rd^2 x \left 
( \p _\mu n_\alpha \right )^2 + i \,\theta \,T \,\,\,,
\label{o3.1}
\ee
where $f$ and $\theta$ are dimensionless coupling constants and 
\be
T\,=\,\frac{1}{8 \pi} \int \rd^2 x \; \epsilon_{\mu \nu}
\epsilon^{\alpha \beta \gamma} n_\alpha \p _\mu n_\beta \p _\nu 
n_\gamma \,\,\, 
\ee
is the integer-valued topological term related to the instanton solutions 
of the model. This model has been the subject of a huge amount of 
study for its theoretical properties and for the large variety of 
its application to condensed matter systems \cite{Haldane,affleck.lectures,ah,zamfat,nersesyan,fendley}
(see also 
\onlinecite{tsvelik}).
For a generic value of $\theta$, the topological term $T$ breaks both the 
$Z_2$ invariance $n_a \rightarrow - n_a$ and the parity space symmetry of 
the action 
${\cal A}_0$, symmetries which are however restored at $\theta = \pi$. 
As a matter of fact, the two values $\theta = (0, \pi)$ are the only ones 
for which the  action (\ref{o3.1}) is known to be integrable. The physical 
properties, however, are completely different in the two cases. At 
$\theta = 0$, the model consists of a $O(3)$ triplet of massive particles, 
with an exact $S$--matrix given in \cite{zam-zam}. At $\theta = \pi$, 
the theory is instead massless \cite{ah,shankar.read,zam-zam.massless} 
and corresponds to the Renormalization Group flow between the $c=2$ CFT 
and the $SU(2)_1$ Wess--Zumino--Witten (WZW) model at level $1$, with central 
charge $c = 1$. In this case the spectrum of the excitations consists of 
massless particles which transform according to the $s=1/2$ representation 
of $SU(2)$, the so--called spinons. The exact massless scattering amplitudes 
for the right and left moving doublets are given in 
\cite{zam-zam.massless}. 

{\em Non--integrable Quantum Field Theories.} One may wonder how 
the spectrum of the theory (\ref{o3.1}) evolves by 
moving the coupling $\theta$, in particular how the two doublets of the 
massless spinons at $\theta = \pi$ are transformed into the triplet of 
massive states at $\theta = 0$. It is of course difficult, if not 
impossible, to provide an exact answer to this question since the model 
is non--integrable for generic values of $\theta$. However, one can gain 
a significant insight about this question by using the Form Factor 
Perturbation Theory (FFPT) proposed in Refs.~\cite{dms,dm}. 
This method allows one to study with a certain accuracy those 
non--integrable models obtained as deformation of an integrable 
quantum field theory. 
For the theory (\ref{o3.1}) we have two possibilities, i.e. 
we can either apply the FFPT in the vicinity of $\theta =0$ 
or use it to analyze the non--integrable theory defined near $\theta = \pi$. 
For reasons that will become clear later, it is simpler to follow the 
evolution of the particle content starting from the value $\theta = \pi$. 
Let us discuss then in more details the model in the vicinity of 
this point. 
  
{\em {Double Sine--Gordon Model}.} At $\theta = \pi$, 
the $O(3)$ sigma model corresponds to a massless flow 
from the CFT with $c=2$ to an infrared fixed point described by the 
$SU(2)_1$ WZW model. In the vicinity of this point, it is appropriate to use 
the conformal fields of the WZW model to write an effective action of the 
sigma model (\ref{o3.1}). This was done in \cite{affleck.lectures,ah} 
and the results can be summarized as follows. Near its infrared fixed point, 
the action ${\cal A}_{\pi}$ corresponds to the $SU(2)_1$ WZW model perturbed by the 
marginally irrelevant perturbation $({\rm Tr} \,g )^2$ (i.e. 
with $\tilde\gamma >0$) 
\be
{\cal A}^{eff}_\pi \,=\, {\cal A}_{SU(2)_1} + \tilde\gamma \int \rd ^2 x 
\,({\rm Tr} \, g )^2 \,\,\,,
\label{o3.2}
\ee
where $g$ is the SU(2) matrix field with conformal dimension 
$\Delta = \overline\Delta = \frac{1}{4}$. Beside the symmetry $SU(2)$, 
this action has also a $Z_2$ invariance related to the transformation 
$g\to -g$. In terms of the WZW fields, the perturbation 
that moves the topological term away from the value $\theta=\pi$ is 
proportional to ${\rm Tr} \,g$ \cite{ah}. This is the only relevant 
field of the WZW model and, moreover, the only $SU(2)$ invariant operator 
in the theory that breaks parity. Thus, for a generic value of 
$\theta$ in the vicinity of $\theta =\pi$, we have an effective 
action given by the following non--integrable perturbation of (\ref{o3.2})  
\be
{\cal A}^{eff} \,=\,{\cal A}_{\pi}^{eff} + 
\tilde \eta \int \rd ^2 x\, {\rm Tr} \, g \,\,\,,  
\label{o3.3}
\ee
with $\tilde\eta \simeq |\theta - \pi|$. At $\tilde
\eta = 0$ the massless particles 
of the theory are the spinons, which can be also viewed as the fundamental 
excitations of the IR point \cite{spinons}. However, the operator 
${\rm Tr} \,g$ is non--local with respect to them. As shown in 
\cite{dm} 
for the case of massive theories, this is the crucial property responsible 
for the confinement of the particles. The same also happens in the massless 
cases \cite{conmus}. Hence, in the presence of ${\rm Tr}\,g$, i.e. 
as soon as we move away from the point $\theta = \pi$, the spinons are 
confined and the model has no longer spin $1/2$ excitations.  
To recover its actual spectrum near the value $\theta = \pi$, it is 
convenient to 
write Eqs. (\ref{o3.2}) and (\ref{o3.3}) in terms of a scalar bosonic field,
$\varphi$, as: ${\cal A}^{eff}_\pi=\int \,d^2x \,\left[ \frac{1}{2}(\partial
\varphi)^2 +\gamma \cos\sqrt{8 \pi}\right]$ and 
$\tilde\eta \, {\rm Tr} \, g =\eta  \cos \sqrt{2\pi} 
\varphi$.
In this new formulation, 
the effective action of the $O(3)$ sigma model in the vicinity of 
$\theta = \pi$ is thus given by the double sine--Gordon model 
\EQ
{\cal A}^{eff} \,=\,\int d^2 x \,\left[
\frac{1}{2} (\partial \varphi)^2 + \eta \,\cos \sqrt{2 \pi}\varphi + 
\gamma \,\cos \sqrt{8\pi} \varphi \,\right]\,\,\,.
\label{doublesinegordon}
\EN 
This is a non--integrable quantum field theory which has been studied 
in details in \cite{dm}. 
In this model, the two periodic interactions play 
a symmetric role and each of the cosine term can be regarded as a 
deformation of the integrable theory defined by the other \cite{footnote1}.

{\em {Affleck's Result.}} Due to the particular values of the cosine frequencies, 
the quantum field theory (\ref{doublesinegordon}) presents a series of remarkable 
peculiarities which, as we are going to show, have far--reaching 
consequences on its spectrum. The first important peculiarity, noticed 
by Affleck \cite{affleck}, is the special pattern of the integrable 
sine--Gordon model at $\beta^2 = 2 \pi$, obtained for  
$\gamma = 0$ in eq.\,(\ref{doublesinegordon}). In fact, the spectrum of this 
integrable model consists of a soliton $s$ and an anti--soliton $\overline s$ 
of mass $m$, which are {\em degenerate} with a breather state $b_1$. Moreover, 
all these particles have the same $S$--matrix. In addition, there is another 
breather state $b_2$ of higher mass, given by $m_2 = \sqrt{3} \,m$. The excitations 
can be then organized into a triplet ($s,\bar s, b_1$) of bosonic states of 
mass $m_t$ and a singlet of mass $m_s = \sqrt{3} m_t$, explicitly 
showing the hidden $SU(2)$ symmetry of the model at this specific 
point \cite{Coleman,haldane.dimerization}. Their exact $S$--matrix 
can be found in \cite{affleck} and is not given here.  

The above pattern for the particles, in particular the triplet of massive 
bosonic states, strongly reminds the spectrum of the original $O(3)$ 
sigma model at $\theta = 0$. However, one may wonder and even doubt 
whether this was just a fortunate coincidence, that would no longer 
persist in presence of the second interaction in the action 
(\ref{doublesinegordon}). The analysis of the double sine--Gordon model shows, 
in fact, that an additional cosine term has generally a drastic impact on 
the spectrum of the solitonic sector of the unperturbed theory producing, in 
particular, their confinement \cite{dm}. However, this circumstance does not occur 
for the theory (\ref{doublesinegordon}) and this is the second remarkable 
peculiarity of the action (\ref{doublesinegordon}). To show that, we 
apply the Form Factor Perturbation Theory.

{\em {Form Factor Perturbation Theory}.}
The Form Factor Perturbation Theory \cite{dms} allows one to estimate 
the variation of the spectrum of an integrable theory, once it 
has been perturbed by an additional term in the action 
$\gamma \int d^2x\, \Psi(x)$. At first order in $\gamma$ one has
\EQ
\delta m^2_i \,\simeq \, 2 \gamma \,F^{\Psi}_{ii}(i \pi)
\,\,\,,
\label{massshift}
\EN 
where $F^{\Psi}_{ii}(\lambda_1 - \lambda_2) = \langle 0 \mid \Psi(0)\mid 
A_i(\lambda_1) \, A_i(\lambda_2)\rangle $ is the two--particle Form Factor 
of the operator $\Psi(x)$ as a function of their rapidities 
parameterizing the dispersion relation $E_i = m_i \cosh\lambda_i$, 
$p_i = m_i \sinh\lambda_i$. For a generic sine--Gordon model 
\EQ
{\cal A} \,=\, \int d^2x \left[\frac{1}{2} (\partial \varphi)^2 + 
g \cos \beta \varphi \right] \,\,\,,
\label{unperturbed}
\EN 
perturbed by another cosine term $\Psi(x) = \cos \alpha \,\varphi(x)$, 
the evaluation of (\ref{massshift}) may be however problematic. 
As shown in 
\cite{dm}, the matrix element of $\cos\alpha\,\varphi(x)$ on 
the soliton states has in general a pole at $\lambda = i\pi$, with a residue 
ruled by the non--locality index of this operator with respect to the 
soliton (The presence of this pole signals the confinement 
of the soliton states in the perturbed theory.). 
Explicitly 
\EQ
-i\,{\mbox Res}_{\lambda = i\pi}F_{s\bar{s}}^\Psi(\lambda) =
\left[1 - \cos(2\pi\alpha/\beta)\right]\,
\langle 0|\cos \alpha\,\varphi(0)|0\rangle\,\,\,.
\label{respsi}
\EN 
However, for the double sine--Gordon (\ref{doublesinegordon}), 
considered as a deformation of the sine--Gordon model 
(\ref{unperturbed}) with $\beta = \sqrt{2\pi}$, we have $\Psi(x) 
= \cos\sqrt{8 \pi} \varphi(x)$, i.e. $\alpha/\beta = 2$, and 
the matrix element $F^{\Psi}_{s\bar{s}}(i\pi)$ is instead finite! 
Moreover, since this Form Factor is determined by the $S$--matrix 
(which is the same for all the particles of the triplet), we have that 
all of them get the same mass correction. In other words, the initial 
triplet identified by Affleck in the theory (\ref{doublesinegordon}) 
at $\gamma =0$ is going to stay degenerate even at $\gamma \neq 0$, a 
result which can be proved to hold at any order in the FFPT. 
It remains, then, to compute its actual correction and to compare it 
with the mass correction of the second breather. Here we 
will only present the basic results of this calculation while their 
complete derivation and the relative discussion will be presented 
somewhere else \cite{conmus}. 
  
The two--particle form factors of the field $\Psi(x) = 
\cos\sqrt{8\pi}\varphi(x)$ on the particles of the triplet and on 
the higher breather $b_2$, can be computed by an analytic 
continuation of the matrix elements of the cluster operators of 
the Sinh--Gordon model \cite{km} (see also 
\cite{lukyanov}). They can be 
written, up to their vacuum normalization, as 
\begin{widetext}

\begin{eqnarray}
F^{\Psi}_{ii}(\lambda) & = & \mu^2 \,
\frac{\sinh^2\frac{1}{2}\lambda}
{\sinh\frac{1}{2}\left(\lambda + i \frac{\pi}{3}\right) 
\sinh\frac{1}{2}\left(\lambda - i \frac{\pi}{3}\right)}
\,\,\frac{1}{{\cal F}(\lambda)} 
\,\,\,\,\,\, ,\,\,\, i=s,\bar s,1\,\,\,; \nonumber \\
F^{\Psi}_{22}(\lambda) & = & -\mu^4 \, 
\frac{1}{6 \sqrt{3} {\cal F}^2(i \frac{\pi}{3})} \, 
\frac{1}{{\cal F}^3(\lambda)} \,\left(1 + \frac{1}{2 
\cosh\frac{1}{2}\left(\lambda + i \frac{\pi}{3}\right) 
\cosh\frac{1}{2}\left(\lambda- i \frac{\pi}{3}\right)}\right)  
\label{importantFF} \\ 
&\times & 
\frac{\sinh^4\frac{1}{2}\lambda}
{
\cosh\frac{1}{2}\left(\lambda + i \frac{2\pi}{3}\right) 
\cosh\frac{1}{2}\left(\lambda - i \frac{2\pi}{3}\right)
\cosh\frac{1}{2}\left(\lambda + i \frac{\pi}{3}\right) 
\cosh\frac{1}{2}\left(\lambda - i \frac{\pi}{3}\right)
}
\,\,
\nonumber 
\end{eqnarray}
where $\mu^2 = \frac{3}{2} \sqrt{3} \,{\cal F}(i\pi)$ and the analytic 
function ${\cal F}(\lambda)$ can be expressed as 
\EQ
{\cal F}(\lambda)\,=\,
\prod_{k=0}^{\infty}
\left|
\frac{\Gamma\left(k+\frac{3}{2}+\frac{i\hat\lambda}{2\pi}\right)
\Gamma\left(k+\frac{2}{3}+\frac{i\hat\lambda}{2\pi}\right)
\Gamma\left(k+\frac{5}{6}+\frac{i\hat\lambda}{2\pi}\right)}
{\Gamma\left(k+\frac{1}{2}+\frac{i\hat\lambda}{2\pi}\right)
\Gamma\left(k+\frac{4}{3} +\frac{i\hat\lambda}{2\pi}\right)
\Gamma\left(k+\frac{7}{6}+\frac{i\hat\lambda}{2\pi}\right)}
\right|^2 \,\,\, , 
\EN
($\hat\lambda = i \pi -\lambda$). This function does not have either 
poles nor zero in the physical strip $0 < {\rm Im}\,\lambda < \pi$ and 
satisfies the functional equations 
\begin{eqnarray*}
{\cal F}(\lambda + i\pi)\, {\cal F}(\lambda) & = &
\frac{\sinh\lambda}{\sinh\lambda + \sinh\frac{i\pi}{3}}\,\,\,; \\
{\cal F}(\lambda + i\frac{\pi}{3})\, {\cal F}(\lambda - i\frac{\pi}{3}) 
& = & 
\frac{\cosh\frac{1}{2}\left(\lambda + i \frac{\pi}{3}\right) 
\cosh\frac{1}{2}\left(\lambda - i \frac{\pi}{3}\right)}
{\sinh^2\frac{\lambda}{2}}\,\,\, {\cal F}(\lambda)  \,\,\,. 
\end{eqnarray*}
\end{widetext}
We have, moreover, the following identity ${\cal F}(i\pi) \,
{\cal F}^2(i \frac{\pi}{3}) = \frac{1}{3}$. Using the expressions 
(\ref{importantFF}) we can now evaluate the mass correction 
(\ref{massshift}), which are given by 
\EQ
\begin{array}{c}
\delta m_t^2 \,= \,2\sqrt{3} \, \gamma \,\,\,; \\
\delta m_s^2 \,= \,6\sqrt{3} \, \gamma \,\,\,, 
\end{array}
\label{differentshift}
\EN
i.e. the first order correction to the mass of the singlet state 
is {\em three times larger} than the one relative to the mass of the 
particles of the triplet (Figure 1). Note that the actual values of 
these corrections depends on the normalization of the operator $\Psi(x)$. 
One can get rid of this problem by considering the universal quantity given by 
their ratio. 

{\begin{figure}[ht]
\centerline{
\fbox{\includegraphics[width=0.40\textwidth]{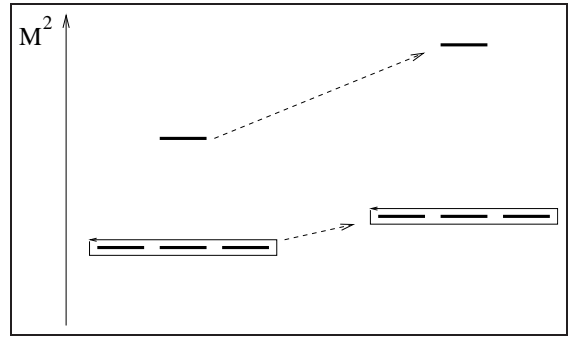}}}
\vspace{3mm}
\caption{Unperturbed and first order correction of the masses 
of the triplet and singlet states.
}
\label{masses}
\end{figure} 
}

This result gives a strong indication that moving away from $\theta =\pi$ 
and going toward the value $\theta =0$, the spectrum will evolve as follows. 
The particles of the triplet remain degenerate and stable also for finite 
values of $\gamma$, alias for all the RG trajectories of the $SU(2)_1$ WZW 
model which asymptotically reach those of the $O(3)$ sigma model with 
$\theta$ in the interval $[0,\pi)$. Hence, these particles are those
which become the triplet of the $O(3)$ sigma model at $\theta =0$ and 
their mass $M_t(\gamma)$ should always be finite. Concerning the singlet, 
the FFPT shows that its mass $M_s(\gamma)$ increases faster than 
$M_t(\gamma)$ by varying $\gamma$. Since at $\theta =0$ there is 
no trace of this state, its mass should become unbounded moving toward 
this value, causing the complete decoupling of this particle from the 
theory. It is then easy to argue, by continuity, that this state corresponds 
to a stable particle of the theory only in an interval of $\theta$ 
near $\theta = \pi$. It becomes instead a resonance below a certain 
critical value $\theta_c$, i.e. above $\gamma_c$ determined by the threshold 
condition $M_s(\gamma_c) \geq 2 M_t(\gamma_c)$. Notice, in particular, 
that at first order in $\gamma$ this equation does not have solution 
for $\gamma > 0$, i.e. the singlet is still a stable particle of the 
theory. The above considerations suggest that the masses should have, 
qualitatively, the behaviour given in Figure 2, with their 
cusp $M_i \simeq (\pi - \theta)^{2/3}$ at $\theta = \pi$ 
dictated by the anomalous dimension of the operator ${\rm Tr}\,g$ \cite{ah}
(logaritmic corrections to the power law behaviour were
considered in  Ref.~\cite{agsz}).

{\begin{figure}[ht]
\centerline{
\fbox{\includegraphics[width=0.40\textwidth]{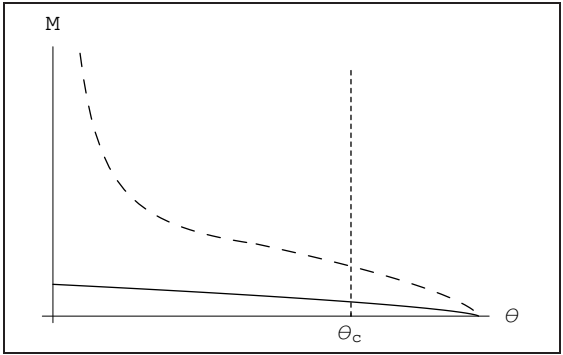}}}
\vspace{3mm}
\caption{
Qualitative behaviour of the masses of the triplet 
(continuum line) and the singlet (long--dashed line) as functions 
of $\theta$ in the interval $(0,\pi)$. The singlet particle is stable 
in the interval above $\theta_c$.
}
\label{curva}
\end{figure} 
}

{\em {Conclusions}.}
The FFPT allowed us to gain new insights on the spectrum 
of the $O(3)$ non--linear sigma model with $\theta$ term in the vicinity of 
$\theta = \pi$. As soon as $\theta$ is moved away from this value, 
the spinons are confined as a consequence of the non--local properties of the 
associated perturbed operator. The spectrum can be obtained by 
analyzing the effective action of the model near its $SU(2)_1$ fixed 
point, given by the double Sine--Gordon model (\ref{doublesinegordon}). 
The pattern of triplet massive states identified by looking at the 
integrable model for $\gamma = 0$ turns out to be robust even in the 
non--integrable field theory with $\gamma \neq 0$. The singlet state, 
on the contrary, belongs to the stable part of the spectrum only in an 
interval of values of $\theta$ close to $\pi$, whereas it becomes a 
resonance below a critical value $\theta_c$. It should not be difficult 
to confirm these predictions by a numerical study of the model and to 
determine correspondingly the critical value $\theta_c$. 

\vspace{3mm}

\begin{acknowledgments} 
We would like to thank G. Delfino for important discussions. We also 
thank I. Affleck for useful comments on the manuscript.
The final part of this work was completed when one of us (G.M.) was 
visiting the Laboratoire de Physique Theorique et Hautes Energies
(LPTHE) in Jussieu, Paris and was attending, later, a conference organised 
in Amsterdam at the Institute for Theoretical Physics (ITP). G.M. 
thanks LPTHE and ITP for the warm hospitality. D.C would also like to 
thank A. Nersesyan, F. Essler and K. Schoutens for discussions. 
D.C. is supported by the European Community under Marie Curie 
Fellowship grant HPMF-CT-2002-01591.
\end{acknowledgments}



\begin{thebibliography}{99}
\bibitem{Haldane} F.D.M. Haldane, {\em Phys. Lett.} {\bf A 93} 
(1983), 464; {\em Phys. Rev. Lett.} {\bf 50} (1983), 1153; 
{\em Journ. Appl. Phys.} {\bf 57} (1985), 3359; I. Affleck, 
{\em Nucl. Phys.} {\bf B 257} (1985), 397.
\bibitem{affleck.lectures} I. Affleck, {\em Quantum Theory Methods 
and Quantum Critical Phenomena}, in {\em Fields, Strings and 
Critical Phenomena}, Les Houches XLIX, 1988.  
\bibitem{ah} I. Affleck and F.D.M. Haldane, {\em Phys. Rev.} 
{\bf B 36} (1987), 5291.
\bibitem{zamfat} V. A. Fateev and Al.B. Zamolodchikov, 
{\em Phys. Lett.} {\bf B 271} (1991), 91; V.A. Fateev, E. Onofri 
and Al.B. Zamolodchikov, {\em Nucl. Phys.} {\bf B 406} (1993), 521.
\bibitem{nersesyan} D.G. Shelton, A.A. Nersesyan and A.M. Tsvelik, 
{\em Phys. Rev.} {\bf B 53} (1996), 8521. 
\bibitem{fendley} P. Fendley, {\em Phys. Rev. Lett.} {\bf 83} (1999), 4468;
{\em Phys. Rev.} {\bf B 63} (2001), 104429; JHEP {\bf 0105} (2001), 050.
\bibitem{tsvelik} A.M. Tsvelik, {\em Quantum Field Theory in Condensed 
Matter Physics}, Cambdridge Univ. Press. 
\bibitem{zam-zam} A.B. Zamolodchikov and Al.B. Zamolodchikov,   
{\em Ann. Phys.} NY {\bf 120} (1979), 253. 
\bibitem{shankar.read} R. Shankar and N. Read, {\em Nucl. Phys.} 
{\bf B 336} (1990), 457.
\bibitem{zam-zam.massless} A.B. Zamolodchikov and Al.B. Zamolodchikov,   
{\em Nucl. Phys.} {\bf B 379} (1992), 602.
\bibitem{dms} G. Delfino, G. Mussardo and P. Simonetti, {\em Nucl. Phys.} 
{\bf B 473} (1996), 469.
\bibitem{dm} G. Delfino and G. Mussardo, {\em Nucl. Phys.} {\bf B 516}, 
(1998), 675.
\bibitem{spinons} P. Bouwknegt, A.W.W. Ludwig and K. Schoutens, 
{\em Phys. Lett.} {\bf B 338} (1994), 448; K. Schoutens, {\em 
Phys. Rev. Lett.} {\bf 79} (1997), 2608; P. Bouwknegt and K. Schoutens, 
{\em Nucl. Phys.} {\bf B 547} (1999), 501. 
\bibitem{conmus} D. Controzzi and G. Mussardo, {\em in preparation}. 
\bibitem{footnote1}
The action (\ref{doublesinegordon}) depends on the dimensionful
coupling $\eta \sim M^{3/2}$, where $M$ is a mass scale, and on the 
(naive) dimensionless coupling constant $\gamma$. However a mass scale 
$M$ is present also in $\gamma$, due to it RG equation. Hence 
the coupling $\eta$ can be used to set the scale for all 
dimensionful quantities of the theory whereas $\gamma(M)$, alias 
$\gamma(\eta^{2/3})$, as a label of the different Renormalization Group 
trajectories which pass close to the fixed point at $\eta = \gamma = 
0$.
\bibitem{affleck} I. Affleck, {\em Nucl. Phys.} {\bf B 265} (1986), 448.
\bibitem{Coleman} S. Coleman, {\em Ann. Phys.} (NY) {\bf 101} (1976), 
239.
\bibitem{haldane.dimerization} F.D.M. Haldane, {\em Phys. Rev.} 
{\bf B 25} (1982), 4925. 
\bibitem{km} A. Koubek and G. Mussardo, {\em Phys. Lett.} {\bf B 311}
(1993), 193.
\bibitem{lukyanov} S. Lukyanov, {\em Mod. Phys. Lett.} {\bf A 12} 
(1997), 2543.
\bibitem{agsz} 
I. Affleck, D. Gepner, H.J. Schulz and T. Ziman, {\em J. Phys. }
{\bf A 22}(1989), 511,
\end{thebibliography}
\end{document}